\documentclass[aps,prl,superscriptaddress,twocolumn,showpacs, letter]{revtex4-1}

\usepackage{amsmath}
\usepackage{graphicx}
\usepackage{color,soul}
\newcommand{\tbmno}{TbMn$_2$O$_5$}

\begin{document}


\title{Antiferromagnetically spin polarized oxygen observed in magneto-electric TbMn$_2$O$_5$}

\author{T.A.W. Beale}
\affiliation{Department of Physics, University of Durham,
Rochester Building, South Road, Durham, DH1 3LE,
UK}
\author{S.B. Wilkins}
\affiliation{Department of Condensed Matter Physics
\& Materials Science, Brookhaven National Laboratory, Upton, New York, 11973-5000, USA}
\author{R.D. Johnson}
\affiliation{Department of Physics, University of Durham,
Rochester Building, South Road, Durham, DH1 3LE,
UK}
\author{S.R. Bland}
\affiliation{Department of Physics, University of Durham,
Rochester Building, South Road, Durham, DH1 3LE,
UK}
\author{Y. Joly}
\affiliation{Institut N\'{e}el, CNRS, and Universit\'{e} Joseph Fourier, B.P. 166, F-38042 Grenoble Cedex 9, France}
\author{T.R.~Forrest}
\author{D.F. McMorrow}
\affiliation{London Centre for Nanotechnology, 17-19 Gordon Street, London, WC1H 0AH, UK}
\author{F. Yakhou}
\affiliation{European Synchrotron Radiation Facility, Bo\^\i te Postal 220, F-38043 Grenoble Cedex, France}
\author{D. Prabhakaran}
\author{A.T. Boothroyd}
\affiliation{Department of Physics, University of Oxford,
Clarendon Laboratory, Parks Road, Oxford, OX1 3PU, UK}
\author{P.D. Hatton}
\affiliation{Department of Physics, University of Durham,
Rochester Building, South Road, Durham, DH1 3LE,
UK}

\date{\today}

\begin{abstract}

We report the direct measurement of antiferromagnetic spin polarization at the oxygen sites in the multiferroic \tbmno, through resonant soft x-ray magnetic scattering.  This supports recent theoretical models suggesting that the oxygen spin polarization is key to the magnetoelectric coupling mechanism.   The spin polarization is observed through a resonantly enhanced diffraction signal at the oxygen $K$ edge at the commensurate antiferromagnetic wavevector. Using the \textsc{fdmnes} code we have accurately reproduced the experimental data.  We have established that the resonance arises through the spin polarization on the oxygen sites hybridized with the square based pyramid Mn$^{3+}$ ions.  Furthermore we have discovered that the position of the Mn$^{3+}$ ion directly influences the oxygen spin polarization.

\end{abstract}


\maketitle

The ability to couple magnetism and ferroelectricity, namely the control of charges by changing the magnetic state of a material, paves the way to the development of novel multifunctional devices.   Recently, the goal of being able to control the ferroelectric state by the application of a magnetic field was realized in TbMnO$_3$\cite{kimura:55}, and launched a new area of research into magnetoelectric multiferroic perovskite materials. In these materials the coupling is provided through complex non-linear magnetic structures. Central to current theories\cite{Katsura:057205,Mostovoy:067601} seeking to explain the spontaneous electric polarization in perovskite manganites, is the presence of a magnetic cycloid, arising from magnetic frustration. Such a magnetic structure breaks global inversion symmetry and allows the development of a ferroelectric moment. It is interesting therefore, that in TbMn$_2$O$_5$, a material found to have a huge magnetoelectric coupling\cite{hur:392}, larger even than that in TbMnO$_3$, the magnetic structure was found to be almost collinear\cite{Chapon:177402}. This suggests that an entirely different mechanism drives the multiferroic behavior in TbMn$_2$O$_5$, opening up a new route for the development of multiferroic devices.   A possible mechanism has been determined by Moskvin and coworkers\cite{moskvin:060102,Moskvin:024102}, who show the importance of the charge transfer between the manganese and oxygen, and the resulting spin polarization of the oxygen sites.

In this letter we report the direct observation of a long range, correlated spin polarization at the oxygen sites in TbMn$_2$O$_5$, observed using resonant soft x-ray scattering at the oxygen $K$ edge. This resonance is seen at the (0.5,0,0.25) antiferromagnetic wavevector corresponding to the magnetic order.  Through \emph{ab-initio} calculations of the incident x-ray energy dependence of the diffraction intensity we show that the ordered spin polarization arises from the oxygen sites hybridized with the Mn$^{3+}$ ions. The temperature dependence of the oxygen spin polarization and the manganese ordering, confirms that the oxygen spin polarization appears at the antiferromagnetic transition, simultaneous to the onset of the ferroelectric dipole moment. We suggest that the oxygen spin polarization is a key factor the mechanism of multiferroicity in TbMn$_2$O$_5$.

\begin{figure}
\includegraphics[width=\columnwidth]{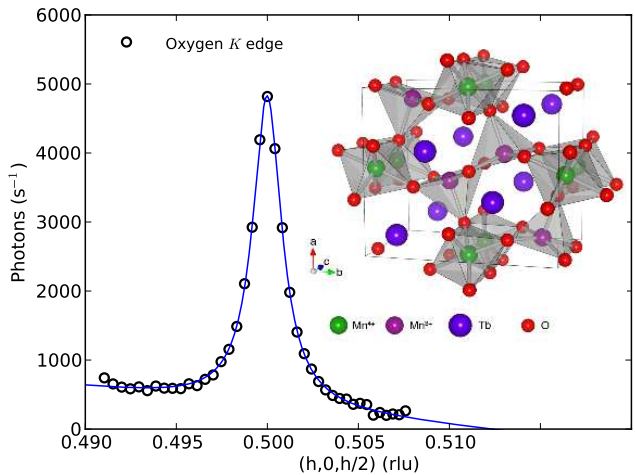}
\caption{$\vec{Q}$ scan of intensity in the [201] direction in reciprocal space at the oxygen $K$ edge at 529.3~eV.  The solid blue line is a fit of a Lorentzian lineshape with a linear background.   The inset represents the $Pbam$ crystal structure of \tbmno, highlighting the two different positions of the manganese ions.   The grey polyhedra show the octahedra centred on the Mn$^{4+}$ ions (green), and the square based pyramids around the Mn$^{3+}$ ions (purple).}
\label{fig:xtal}
\end{figure}

\tbmno\ crystallizes into the orthorhombic $Pbam$ spacegroup\cite{alonso:8515}, with the Mn$^{3+}$ and Mn$^{4+}$ ions occupying Wyckoff $4h$ and $4f$ positions respectively\cite{alonso:8515} (Fig.~\ref{fig:xtal}).   This places the Mn$^{3+}$ ions in the square based pyramids, and the Mn$^{4+}$ ions at the centre of the octahedra.  A commensurate magnetic structure is observed between 24 and 33~K, where both Mn$^{3+}$ and Mn$^{4+}$ ions form antiferromagnetic chains in the $ab$ plane alternately coupling ferro- and antiferro-magnetically ($\uparrow\uparrow\downarrow\downarrow$) along the $c$-axis\cite{Chapon:177402, blake:214402}.     Incommensurate magnetic phases occur above and below this commensurate phase, and the system is paramagnetic above 43~K.   Below 10~K the rare earth ions order antiferromagnetically, although recent resonant x-ray studies have shown a spin polarized terbium $5d$ band in all magnetically ordered phases\cite{ewings:104415,johnson}.   At 38~K, a large peak in the dielectric constant is observed\cite{hur:392}, due to the formation of a spontaneous ferroelectric polarization along the $b$-axis.    A dramatic drop in the electric polarization is observed upon further cooling below 24~K, concomitant with the low temperature commensurate to incommensurate magnetic transition.

\begin{figure}
\includegraphics[width=\columnwidth]{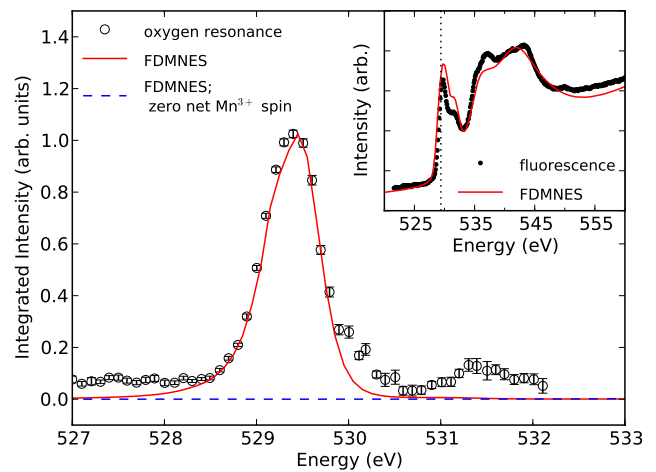}
\caption{Integrated intensity of $\vec{Q}$  scans in the $[201]$ direction through the (0.5,0,0.25) reflection, at energies from 527~eV to 532~eV through the oxygen $K$ edge.  (\emph{Inset})  Fluorescence measurement at the oxygen $K$ absorption edge, with a dotted line indicating the peak of the resonant signal.  The solid red lines in both panels display the \textsc{fdmnes} calculations, the dashed blue line in the main panel shows the result from a simulation with zero net spin on the Mn$^{3+}$ ions.}
\label{fig:ox}
\end{figure}

Soft x-ray scattering provides unique insights into the physics of strongly correlated transition metal oxides, and multiferroic materials\cite{Okamoto:157202,Wilkins:207602}, partly due to the great advantage of performing an element specific magnetic measurement.   By tuning to a particular absorption edge it is possible to resonantly enhance scattering from specific ions in a material\cite{Beale:174432}. Our scattering experiments were undertaken at ID08, ESRF and 5U.1, SRS Daresbury. A single crystal sample of \tbmno\ was prepared though pre-alignement and polishing such that the (201) wavevector was surface normal.  Figure~\ref{fig:xtal} shows the (0.5,0,0.25) antiferromagnetic signal at the oxygen $K$ edge observed upon cooling below 38~K. Figure~\ref{fig:ox} shows the incident photon energy dependence of the integrated intensity of the (0.5,0,0.25) wavevector, and the fluorescence yield (inset), in the vicinity of the oxygen $K$ absorption edge.  Such strong resonant enhancements were also observed at the manganese $L_{2,3}$ and terbium $M_{4,5}$ edges.  The resonant data in Fig.~\ref{fig:ox} was collected through a series of $\vec{Q}$ scans across the energy range removing the fluorescence background. The $\vec{Q}$ scan of the diffraction peak (Fig.~\ref{fig:xtal}) shows that the peak width is limited by the attenuation of the beam by the sample, providing a lower bound on the correlation length of $\sim2000$~\AA.  The intensity of the reflection was $5000$~s$^{-1}$, compared to $400,000$~s$^{-1}$ at the manganese $L_3$ edge.   

This surprising observation of a diffraction signal at the oxygen $K$ edge indicates an antiferromagnetically ordered spin polarization on the oxygen sites.

\begin{figure}
\includegraphics[width=\columnwidth]{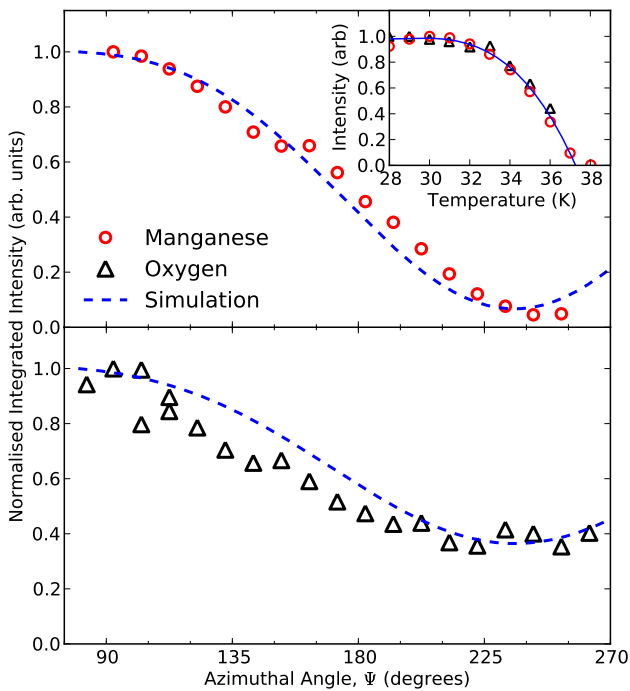}
\caption{Variation of intensity of the diffraction signal at the manganese $L_3$ and oxygen $K$ edges with azimuthal rotation of the sample around the scattering vector, with vertically ($\pi$) polarized incident x-rays.  The dashed lines show simulations of the azimuthal dependence as calculated from the manganese moments determined by Blake \emph{et al.}\cite{blake:214402}.   The fitting errors for the azimuth are within the size of the data points.  (\emph{inset})  Temperature dependence of the (0.5,0,0.25) reflection at the manganese $L_3$ edge (644.1~eV) and oxygen $K$ edge (529.4~eV).  The solid line shows a guide to the eye.}
\label{fig:azi}
\end{figure}

To confirm that the scattering observed at the oxygen $K$ edge is indeed magnetic, we measured the azimuthal dependence, at both the oxygen $K$ edge and manganese $L_3$ edge.  The resonant x-ray scattering cross-section is dependent on both the incident x-ray energy, the polarization of the incident and outgoing x-rays, and the Fourier component of the magnetic structure. The energy dependence gives us the resonant profile as shown in Fig.~\ref{fig:ox}, while the polarization dependence indicates the direction of the Fourier component.   It is possible to take a measurement sensitive to such a direction by rotating the sample around the scattering vector, thereby changing the projection of the electric field of the incident x-rays with the magnetic moment.  Figure~\ref{fig:azi} shows the azimuthal dependence measured with vertically ($\pi$) polarized incident x-rays at the manganese ${L_3}$ edge and oxygen $K$ edge sensitive to the direction of the Fourier components on the manganese and oxygen sites respectively.  The dashed lines superimposed on these data are calculations based on the magnetic structure as reported by Blake\cite{blake:214402}, and the resonant x-ray cross-section\cite{Hill:236}.  The oxygen $K$ and manganese $L_3$ edge data shows exactly the same anisotropy with the differences between the curves arising solely due to geometrical differences due to the different photon energies. It is apparent therefore, that both the spin polarized oxygen and the manganese possess a common magnetic structure.

The temperature dependence of the magnetic reflection shown in the inset of Fig.~\ref{fig:azi}, displays a transition at both the oxygen and manganese edges at 38~K.  The identical temperature dependancies further confirms the link between the spin polarization on the oxygen and the magnetic structure on the manganese ions.  There has been some confusion over the coexistence of the magnetic and electric polarization transitions.    Neutron diffraction studies have reported a magnetic transition at 33~K\cite{Chapon:177402}, some 5~K lower than the electric polarization transition.   This has caused difficulty for the analysis of transitions between the magnetic point groups of incommensurate and commensurate magnetic structures\cite{Toledano:144103}.    Our data, in agreement with earlier x-ray\cite{Koo:197601} and neutron diffraction studies\cite{Koo:562},  strongly suggests a simultaneous magnetic and ferroelectric transition, simplifying the phase diagram.   Above 38~K we observed no intensity in the high temperature incommensurate phase at the expected wavevector, in contrast to ErMn$_2$O$_5$ in which a smooth transition was seen between the commensurate and incommensurate state\cite{bodenthin:027201}.  This separation of the commensurate and incommensurate reflections, has been observed in an earlier soft x-ray diffraction study\cite{Okamoto:157202}, and also through neutron diffraction\cite{kobayashi:3439}.

To establish the origin of the spin polarization we performed calculations with the \textsc{fdmnes} code\cite{joly:125120}, to model the observed resonance profile.  \textsc{fdmnes} is an \emph{ab-inito} cluster based mono-electronic code, written for calculated x-ray absorption and x-ray resonant diffraction signals.   Figure~\ref{fig:ox} shows the results of the \textsc{fdmnes} calculations based on the crystal structure of Blake~\emph{et al}.\cite{blake:214402}, with the oxygen positions corresponding to the room temperature neutron diffraction data\cite{alonso:8515}, the only data currently available.  The calculation used a magnetic unit cell of size $2a\times b\times4c$ containing 160 oxygen atoms at 40 inequivalent sites, calculated over a cluster radius of $4~\mathrm{\AA}$, with an average of 23 atoms per cluster.    The simulated intensity arose entirely from the E1-E1 electric dipole terms.   A successful fit to the data  is achieved using the \textsc{fdmnes} code, with a Fermi energy of 529.6~eV, coinciding with the resonant signal. The simulation of the fluorescence yield, based upon absorption calculations, is in excellent agreement with the experimentally measured curve as shown in the inset of Fig.~\ref{fig:ox}. The \textsc{fdmnes} simulation of the resonance not only correctly predicts the energy corresponding to the maximum intensity, but also reproduces the main spectral feature. 

\begin{figure}
\includegraphics[width=\columnwidth]{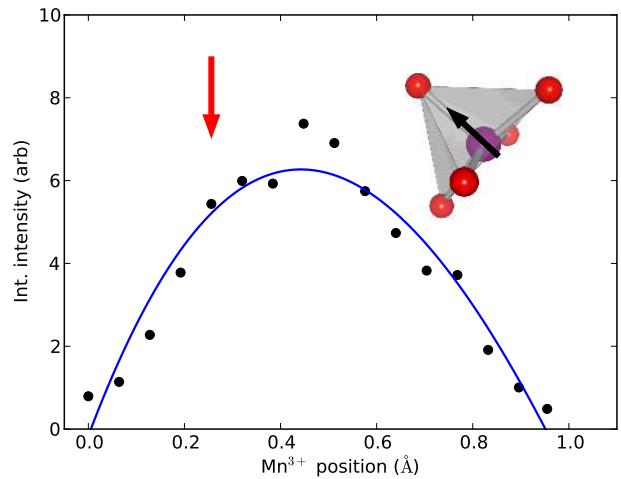}
\caption{The integrated intensity of the calculated resonance through varying the Mn$^{3+}$ position toward the apical oxygen, out of the basel plane (schematic).  The solid line is a guide to the eye, and the vertical red arrow indicates the empirically determined position of the Mn$^{3+}$ ion\cite{blake:214402}.}
\label{fig:dos}
\end{figure}

Calculation of the density of states of the manganese ions show that at the energy corresponding to the oxygen $K$ edge resonance, the only available final states correspond to the Mn$^{3+}$ ions. By contrast the nearest available Mn$^{4+}$ states are $\sim$10~eV higher in energy, suggesting that the resonant signal observed at the oxygen $K$ edge arises purely from long range order of the spin polarization of the oxygen sites through the  hybridization with the Mn$^{3+}$ ions.  A second \textsc{fdmnes} simulation was performed with zero net spin on the Mn$^{3+}$ ion, shown by a dashed line in Fig.\ref{fig:ox}. This shows no calculated intensity at the (0.5,0,0.25) wavevector confirming this hypothesis.

The $Pbam$ symmetry of \tbmno\ allows freedom of the position of the Mn$^{3+}$ ion along the axis of the pyramid.    Such movement alters the Mn-O bonds, and thus the hybridization between the manganese $3d$ and oxygen $2p$ states.   We investigated the effect of the position of the Mn$^{3+}$ within the square based pyramid, and how this affects the antiferromagnetic spin polarization of the oxygen sites, through calculating the intensity of the (0.5,0,0.25) wavevector at the oxygen $K$ edge. Figure~\ref{fig:dos} shows how the simulated resonant scattering intensity varies as a function of the position of the Mn$^{3+}$ ion as it is moved out of the basal plane of the pyramid toward the apical oxygen.  The zero position corresponds to the manganese ion sitting in the basal plane, and the arrow shows the experimentally determined position of the Mn$^{3+}$in TbMn$_2$O$_5$\cite{blake:214402}. The structure factor corrected result shows that as the Mn$^{3+}$ ion is moved out of the basal plane, the degree of antiferromagnetic spin polarization increases, shown by the increasing intensity of the (0.5,0,0.25) signal.   The calculation indicates that when the manganese is in the basal plane, and thus the Mn-O-Mn bonds are parallel, there is minimal long range ordered spin polarization with the magnetic wavevector.  However, the antiferromagnetically ordered spin polarization is close to the maximum when the Mn$^{3+}$ ions are at the empirically measured position.

Two theories currently compete to explain the origin of the spontaneous polarization in \tbmno.  The first, ``ionic polarization'' model, requires a displacement of ions from their centrosymmetric positions thereby generating an electric polarization.   This ionic displacement is extremely small, with calculated displacements in the order of $10^{-3}$\AA\cite{Wang:134113}, and has not yet been measured, although an experimental method has been suggested\cite{azimonte:012103}.   In the second model; the spin configuration induces a charge redistribution around the ion nucleus, thus supporting an electric polarization.    This model is supported by optical second harmonic generation results\cite{Lottermoser:100101}, and requires a strong Mn-O hybridization and spin polarization of the oxygen ligands\cite{moskvin:060102}.

A strong electric polarization is produced from this parity breaking exchange mechanism where the electric polarization on the oxygen ($\mathbf{P}_\mathrm{O}$) can be written\cite{Moskvin:024102}:

\begin{equation}
\mathbf{P}_{\mathrm{O}}=\sum_n\Pi_n(\langle\mathbf{S}_\mathrm{O}\rangle\cdot\mathbf{S}_n)
\end{equation}
where $\Pi_n$ is the effective dipole moment of the manganese orbital states, $\mathbf{S}_n$ are the spins of the manganese, and $\langle\mathbf{S}_\mathrm{O}\rangle$ is the spin polarization of the oxygen ligands.   It is reasonable to assume that $\langle\mathbf{S}_\mathrm{O}\rangle\cdot\mathbf{S}_n$ is maximised by the common structure of the ordered spin moment on the manganese and the oxygen spin polarization, demonstrated here through the common $\vec{Q}$ and azimuthal dependencies.   Our data, clearly showing a long range order of the spin polarization of the oxygen sites induced by the hybridization with the Mn$^{3+}$ ions, reinforces previous work showing that the MnO$_5$ pyramid contributes particularly strongly to the electric polarization\cite{moskvin:060102}.

In conclusion the presence of a resonantly enhanced soft x-ray diffraction signal at the oxygen $K$ edge at the (0.5,0,0.25) wavevector proves the existence of a significant, long range correlated order of the spin polarization of the oxygen sites.  The temperature dependence reveals a common transition temperature, coinciding with the commensurate transition, confirming the  magnetic origin of both reflections.   Comparing the azimuthal dependencies of the signals observed at the manganese $L_3$ edge and oxygen $K$ edge we establish a common magnetic structure for both elemental sites, that of the previously refined manganese magnetic structure. Through simulating the resonance using the \textsc{fdmnes} code we have shown that the resonance arises from the order of the spin polarization on the oxygen sites hybridized by the Mn$^{3+}$ ion.   Such a spin polarization of the oxygen is a crucial component in the model of Moskvin and Pisarev\cite{moskvin:060102} to explain the magneto-electric coupling in \tbmno.   Furthermore we show that the position of the manganese within the square based pyramid in \tbmno\ is such that the ordered spin polarization is dramatically increased when compared to the basal plane ion position.   We believe that further studies of oxygen spin polarization in multiferroics will lead to great insights in the understanding of the microscopic mechanism in multiferroic materials.

The authors would like to thank John Hill for stimulating discussions. The image shown in Fig.~\ref{fig:xtal} was depicted using \textsc{vesta}\cite{vesta}. TAWB, SRB, RDJ and TRF would like to acknowledge support from EPSRC and STFC.   The work at Brookhaven National Laboratory is supported by the Office of Science, U.S. Department of Energy, under contract no. DE-AC02-98CH10886.

\bibliography{tbmn2o5}

\end{document}